\documentstyle[aps,prl,twocolumn]{revtex}
\begin{document}
\twocolumn[\hsize\textwidth\columnwidth\hsize\csname@twocolumnfalse\endcsname
\title
{Training a perceptron in a discrete weight space}
\author
{ Michal Rosen--Zvi and Ido Kanter }

   \noindent

\address{ 
    Minerva Center and the Department of Physics, Bar--Ilan University\\
   \mbox{~} Ramat--Gan, 52900, Israel \\
}

\maketitle
\begin{abstract}
On-line and batch learning of a perceptron in a discrete 
weight space, where each weight can take $2 L+1$ different values,
 are examined analytically and numerically. 
The learning algorithm is based on the training of the continuous 
perceptron and prediction following the clipped weights. 
The learning is described by a new set of order parameters, 
composed of the overlaps between the teacher and the continuous/clipped 
students.
Different scenarios are examined among them on-line learning 
with  discrete/continuous transfer functions and  off-line Hebb learning.
The generalization error of the clipped weights decays asymptotically
as $exp(-K \alpha^2)$/$exp(-e^{|\lambda| \alpha})$ in the case of  on-line 
learning with binary/continuous activation functions, respectively, where
$\alpha$ is  the number of examples divided by N, the size of the input 
vector and $K$ is a positive constant that decays linearly with $1/L$.  
For finite $N$ and $L$, a perfect agreement between the discrete student and 
the teacher is obtained for $\alpha \propto \sqrt{L \ln(NL)}$.
A crossover to the generalization error $\propto 1/\alpha$, characterized
continuous weights with binary output, is obtained for
synaptic depth $L > O(\sqrt{N})$.

\end{abstract}
\pacs{} ]
\section{Introduction}
Neural networks and the perceptron as the simplest prototype have become 
most popular as a tool for understanding human learning and as a basis 
for many various applications \cite{Herz,WatkinRauBiehl}.
We are interested in the perceptron learning ability as an archetype
for machines that are able to learn.
Most of the perceptrons that have been studied until now are under two 
totally different  constraints, two extremes. 
Either the teacher weight vector is restricted to a  binary space, 
(the Ising teacher), or it is continuous, confined to a hypersphere.
Only a few aspects of the learning ability of weights which are 
confined to have a finite  number of values have been studied,
although the realistic case on digital computers where numbers
have a finite depth representation is described by this model.
Furthermore, the applicability of neural networks to biology and to the 
construction of real devices requires the understanding of the interplay 
between the weights depth and the  network ability of learning.
Those systems are the intermediate case, 
in which the weights are confined to finite 
space, $(2L+1)^N$, when $L$ is an integer and $N$ 
stands for the input size.
\cite{GutfreundStein,Kanter,MeirFontanary}.

The generalization ability of such networks, in which the synapse has
a finite depth has been analyzed by using replica calculations and 
has been found to have interesting nontrivial behavior of phase 
transition. The learning procedure composed of two phases; one in which the 
learning ability is very limited,  the generalization error is finite, another
phase is when the generalization error is exactly zero, perfect learning is 
gained and it happens in a finite $\alpha$, where $\alpha$ is the 
number of patterns divided by the size of the input  $N$, \cite{MeirFontanary}.
Nevertheless, replica calculations do not involve practical algorithms 
that one may use in order to obtain that learning behavior.
In the Ising case, for instance, although a phase transition 
was predicted, no practical algorithm reproduces this discontinuous 
behavior \cite{SchietseBoutenBroeck,BoutenReimers}.

In contrast to the batch learning, when all the examples are used
together to achieve perfect learning, on-line learning is a procedure
in which an update rule is used and  learning in each step utilizes only  
the last of a sequence of examples. 
Such an algorithm  drastically reduces the computational effort compared 
with batch learning and no explicit storage of a training set is required
\cite{PercBiehl}.
It was shown that there is no updating rule that uses only the discrete 
vector for updating and results in perfect learning \cite{KinzelUrbanczik}.

In this paper we address the issue of practical learning from a finite depth
teacher. The method we introduce is based on the clipping of a continuous 
perceptron. Having an artificial continuous weight vector enables smooth 
learning; clipping it results in a discrete student $\vec{W}^S$, 
whose components are close to those of the teacher.
This method has been used successfully in the Ising perceptron 
\cite{SchietseBoutenBroeck,BoutenReimers,Clipped,Mi}. 
The questions that arise from the procedure above are; whether learning is 
possible at all and if it is possible, does it give better results then 
the learning in a continuous space.
It seems very natural that if the weights'' depth is very large, i.e. there are
many possible values to each weight, the learning behavior of the discrete
weights will be exactly the same as those of a continuous weight.
However, in the following we examine if and what are the scaling relations
between both properties, $L$ and $N$. 

Our main results are:
(a) Learning in the case of finite depth is possible by using a continuous
precursor. This result was confirmed both analytically and numerically.
(b) In the on-line learning scenario: 
Having a binary output results in a fast decay of the generalization
error and at the large $\alpha$ regime 
it decays super-exponentially with $\alpha$, 
$\epsilon_g \propto \exp(-K\alpha^2)$ where  $K$ is some constant. 
Having a continuous output results in a much fast decay of
the generalization error,$\exp(-K_1\exp(K_2\alpha))$,
where $K_i$ are positive constants.
(c) In batch Hebbian learning, 
having a binary activation function, the generalization error 
falls exponentially with $\alpha$.
(d) Perfect learning is obtained when $N$ is very large but finite, 
unlike the continuous perceptrons performance.
Quantitatively, for a given $N$ and $L$ the 
perfect learning is achieved  for $\alpha_f \propto 
O(\sqrt{L \ln( L N)})$. 
(e) A crossover to the behavior of the generalization error in the 
presence of continuous weights occurs for 
$L> o(\sqrt{N})$.

The paper is organized as follows: In section II the architectures and the 
dynamical rules are defined as well as the continuous and discrete students.
In section III the order parameters are defined  and the relations
between the overlaps of the continuous teacher with the discrete/continuous 
students are derived analytically. In section IV, the dynamical evolution of 
the order parameters in the case of binary output  is derived analytically 
and confirmed by simulations. Both,  on-line scenario and Hebbian 
learning are examined. 
In section V the case of large synaptic depth and the crossover to the 
continuous weights is studied. 
In section VI, the perfect learning in finite $N$ systems is examined 
both analytically and numerically. 
Section VII is devoted to analyze results in the case of continuous output.
Finally, in section VIII results are concluded and 
open questions are addressed.

\section{The Model}

\subsection{The Architecture}

We investigate a teacher-student scenario where  both  nets are 
single-layer feed-forward. The examples are generated by the so-called
teacher, which is known to be restricted to a well-defined discrete set of 
values.
We define a synaptic depth L and a set of digital values to be as 
follows  \cite{GutfreundStein,MeirFontanary},
\begin{equation}
W_i^T= \pm \frac{1}{L},\pm \frac{2}{L}... \pm1 .
\label{nozero}
\end{equation}
In case that the zero value is part of the game, the possible values of the 
weights are  
\begin{equation}
W_i^T= 0, \pm \frac{1}{L},\pm \frac{2}{L}... \pm1 .
\label{zero}
\end{equation}
For the sake of simplicity, we present results in this paper
 only for the including zero case (Eq. \ref{zero}).
It is easy to generalize our results for the other case, (Eq. \ref{nozero}).   

The input patterns $\vec{\xi}^\mu$ are chosen at random and independent
of each other. In the following they are drawn from a 
Gaussian distribution with zero mean and unit variance.
The size of the teacher, the student and the input is N.
For any input $\vec{\xi}$ the so-called teacher generates an  output, S, 
according to some rule 
\begin{equation}
S=F(\frac{\vec{W}^T\cdot\vec{\xi}}{\sqrt{N}}). 
\label{rule}
\end{equation}
In the following we will discuss both binary and continuous rules. 
The student has in mind the rule F and the discrete set of values that the 
teacher is confined to.
In addition, in an on-line learning scenario, the student is given in each 
time step, $\mu$, the input $\vec{\xi}^\mu$ and the teacher's output $S^\mu$,
whereas in batch learning the set $(\vec{\xi}^\mu,S^\mu)$  $\mu=1...\alpha N$
is given altogether.

\subsection{Dynamics of the Weights}

A continuous precursor for the student, $\vec{J}$ is needed for 
learning from a discrete teacher. The learning procedure, having a continuous
student, is well known.
In an on-line scenario at each step the continuous student updates its weight 
vector according to some learning algorithm (f).
The generic form of the learning algorithm is
\begin{equation}
\vec{J}^{\mu+1}
=\vec{J}^{\mu}+\frac{\eta}{\sqrt{N}}
f(S^\mu,x_J^\mu)\vec{\xi}^\mu S^\mu,
\label{update}
\end{equation}
where $\eta$ is the learning rate and $x_J$ is the student's local field,  
$x_J \equiv \frac{1}{\sqrt{N}} \vec{J}\cdot\vec{\xi}$.
Such a learning algorithm  means that at each learning step $\mu$, 
the current weight vector $\vec{J}^{\mu}$ is updated according to 
the new example, $\vec{\xi}^\mu$ and each example is presented only once.

In an off-line scenario, there is a set of examples $\vec{\xi}^\mu$ 
$\mu=1...\alpha N$ and they are used altogether to gain perfect learning.
There are methods in which the off-line leaning is made according to a rule 
that defines an additive quantity of all the examples.
The Hebb learning is an archetype of those methods,
\begin{equation}
\vec{J}^{Hebb}
=\sum_{\mu=1}^{\alpha N} \vec{\xi}^\mu S^\mu.
\label{Hebb}
\end{equation}
Such procedures were shown to end up in perfect learning  
\cite{Vallet,ValletGailton}.
Since having a discrete teacher is merely a special case,
not using the knowledge  that the teacher is confined to a discrete 
set of values gives the well-known results; an exponential decay in the case
of continuous rule (on-line learning \cite{SollaSaad,RosenZviBiehlKanter}) and
a power law decay in the case of binary rule (on-line and off-line learning
\cite{Vallet,ValletGailton,Opper,Caticha,BiehlSchwarze}).
 
The way to gain from the knowledge of the discrete nature of the 
weights is in the center of our work, and it is based on having in 
addition a discrete student $\vec{W}^S$ derived from the continuous one 
using the following clipping procedure.
A continuous weight is clipped to the nearest discrete value, among
the $2 L +1$ possibilities. 
Such a clipping  procedure is the optimal one with  the lack
of any prior knowledge about the weights accept that each value 
appears with the same probability.
We define limit values, $\lambda_\l$, which are arranged in increasing order.
The limit values  divide the continuous region of the precursor weight 
vector components  to $2 L +1$ intervals, according to the number 
of the available values as in  Eq. \ref{zero}.
The clipping process is such that $J_i$ is mapped onto $\frac{\l}{L}$ for
 $J_i \in (\lambda_{\l}$, $\lambda_\l+1)$.
The set of limits includes $\{\lambda_{-\l},\lambda_{-\l-1}, ...\lambda_{-1},
\lambda_{0}, \lambda_{1}... \lambda_{\l+1}\}$. 
It is given by the following mathematical rule:
\begin{equation}
W_i^S =\sum_{\l=-L} ^{L} \frac{\l}{L}
[ \theta(\lambda_{\l+1}-J_i)-\theta(\lambda_{\l}-J_i)]
\label{map}
\end{equation}
where  $\theta$ is the Heavyside function.
  
Since the value of those limits, $\lambda_\l$, is somewhat unclear, 
we would like to exemplify it with some specific cases.
In the case of $L=1$ ,Eq. \ref{nozero},
for instance, due to symmetry it is obvious that the limit between -1 and 1 
should be 0. Hence, one introduces the following limits, 
$\lambda_{-1}= -\infty$, $\lambda_{0}= 0$, $\lambda_{1}= \infty$. 
Evaluating the mapping equation results in the well known clipping rule, 
$W_i^S=sign(J_i)$, \cite{SchietseBoutenBroeck,Clipped}.
Finding the appropriate value for all other cases but the Ising perceptron
becomes more complicated, the continuous space is no longer 
divided  into two clear regions
and hence one has to consider carefully the value of the limits.  

In this paper we chose to nail down the general results by focusing in the 
including zero case, $L=1$, i.e., $W_i= 0,\pm 1$.
This case is known as the diluted Ising case  and some other 
aspects of it  have been studied in references
\cite{BoutenKomoda,GutfruendStein,Malzahn}.
It contains the simplicity of the Ising case on the one hand and introduces a
more generality concerning digital values on the other hand. 
In this case, there is only one unknown parameter, $\lambda_{1}$, since 
$\lambda_{2}= -\lambda_{-1}=\infty$, and  $\lambda_{0}= -\lambda_{1}$.

While choosing the value of the limits, (in the last case that means only 
choosing the value of $\lambda_{1}$) one should take into consideration the 
a priori knowledge about the weights of teacher.
It is clear that the limits should scale with the student norm, since the 
exact set of values that the continuous student end up with is irrelevant.
The mapping rule ensures that the digital student ends up with the same 
values as those of the teacher.
This will be shown only after analyzing the new order parameters and their 
dependent on the former one, as is presented in the next chapter.

\section{The Order Parameters}

Evaluating the agreement between teacher and student is made by calculating
either the generalization error or the order parameters.
The generalization error, $\epsilon_g$, is calculated by taking the average 
of the student/teacher disagreement over the distribution of input vectors. 
The generalization error is given, in principle, by the overlaps between 
the vectors, (the so-called order parameters).
However, in order to get into details one has first to define the rule, (
F in Eq. \ref{rule}).
This will be done in the next sections.
In the following we concentrate in introducing the complete set of order 
parameters and their inter-relations.

In our case there are three vectors and hence two interdependent  
sets of order parameters; one set concerns the continuous overlaps,
\begin{eqnarray}
R_J
\equiv \frac{1}{N}
\vec{J}\cdot\vec{W}^T,\nonumber \\
Q_J
\equiv \frac{1}{N}
\vec{J}\cdot\vec{J},\nonumber \\
\label{ContOrd}
\end{eqnarray}
and one set concerns the digital vector's overlaps,
\begin{eqnarray}
R_W
\equiv \frac{1}{N}
\vec{W}^S\cdot\vec{W}^T,\nonumber \\ 
Q_W
\equiv \frac{1}{N}
\vec{W}^S\cdot\vec{W}^S.\nonumber \\
\label{DiscOrd}
\end{eqnarray}
Note that the dynamical evolution of the continuous set of order parameters,
Eq. \ref{ContOrd}, is independent of the clipped order parameters, since 
the {\it training} is done only following the continuous weights.
In contrary to the training process the {\it prediction} and the 
generalization error is made following the clipped student.
Hence, finding the quantitative interplay between the continuous set of
order parameters, Eq. \ref{ContOrd}, and the discrete set of
order parameters, Eq. \ref{DiscOrd}, is the cornerstone for the analytical
description of the generalization ability of the student.

In this section we examine the relation between the clipped set 
and the continuous one. The development of $R_J$, $Q_J$
is not influenced by the clipping method. Hence, finding out  the above
relation enables finding the development of the clipped order parameters
and results in a description that gives the whole picture of the learning  
process.  

The teacher's norm is determined according to the a-priori probabilities for 
each discrete value. Having equal probability and taking the thermodynamic
limit results in the norm,
\begin{equation}
T
\equiv \frac{1}{N}
\vec{W}^T\cdot\vec{W}^T=
\frac{1}{L^2 n_L} \sum_{\l=1}^{L} \l^2
=\frac{1}{3}+\frac{1}{3 L},
\end{equation}
where $n_L$ defined to be the number of optional values, $n_L=2L+1$. 
The order parameters in the clipped machines, $R_W$ and  $Q_W$,  
as a function of those of the continuous machine, 
$R_J$ and $Q_J$, are evaluated as follow:
\begin{eqnarray}
R_W =
< \frac{1} {N}
\sum W_i^T \frac{\l} {L} 
[\theta(\lambda_{\l+1}-J_i)-\theta(\lambda_{\l}-J_i)]>\nonumber \\
Q_W =
<\frac{1}{N}
\sum \frac{\l^2}{L^2}
[\theta(\lambda_{\l+1}-J_i)-\theta(\lambda_{\l}-J_i)]^2>\nonumber \\
\label{RQ}
\end{eqnarray}
where $<A>$ is an average over the known constraints and the known overlaps, 
\begin{equation}
<A> \equiv 
\frac{Tr_{W^T} \int  dJ_i  \delta( J_i^2-NQ_J) \delta(J_i W_i^T-NR_J)A}
{Tr_{W^T} \int  dJ_i  \delta( J_i^2-NQ_J) \delta(J_i W_i^T-NR_J)}.
\end{equation}  
The validity of this average is based on the assumption that 
all vectors $\vec{J}$ which are consistent with the constraints are taken
with equal probability.
This assumption is violated in case that  the updating of the continuous
vector itself is made according to the clipped one,
 see \cite{SchietseBoutenBroeck,Mi}.

The results  are: 
\begin{eqnarray}
R_W =
\frac{1}{2 L^2 n_L}
\sum \l \l'
[erf(\Phi_{\l+1,\l'})
-erf(\Phi_{\l,\l'})], \nonumber \\
Q_W =
\frac{1}{2 L^2 n_L}
\sum \l^2
[erf(\Phi_{\l+1,\l'})
-erf(\Phi_{\l,\l'})] ,\nonumber \\
\label{Map}
\end{eqnarray}
where the summation  is over all the possible values, 
starting from $\l,\l'=-L,~-L+1, ...,L$ and we defined
\begin{equation} 
\Phi_{\l,\l'} \equiv
\frac{ \frac{\lambda_\l}{\sqrt{Q_J}}-\frac{\rho_J}{\sqrt{T}} \frac{\l'}{L}}
{\sqrt{2(1-\rho_J^2)}}),
\label{Phi}
\end{equation}
where $\rho_J \equiv \frac{R_J}{ \sqrt{T}\sqrt{Q_J}}$, $\rho_W \equiv \frac{R_W}{ \sqrt{T}\sqrt{Q_W}}$ are the geometrical order parameters.

In the limit $L \rightarrow \infty$ the summation in Eq. \ref{Map}
 can be replaced by an integral.
Calculating the integrals in this  limit results in the
obvious identities, $R_W=R_J$ and $Q_W=Q_J$.
Note that taking integrals instead of summation imposes an inequality.
The difference $\Phi_{\l,\l'}-\Phi_{\l+1,\l'}$ tends to zero as 
long as  $L>>1/\sqrt{1-\rho_J^2}$, (see Eq. \ref{Phi}).
Hence, in the event that $L$ is very large, 
learning with the continuous student or learning with the clipped version 
performs the same result as long as  $\rho_J$ is smaller then $2/L$. 
This limit is discussed in section VI.

We exemplify the general results in the case of the diluted Ising perceptron.
In that case we used the following limits, 
\begin{eqnarray}
\lambda_{2}=-\lambda_{-1}=\infty \nonumber \\
\lambda_{1}=-\lambda_{0} \nonumber \\
\label{limits}
\end{eqnarray}
and the teacher's norm is $T=2/3$. The mapping above gives
\begin{eqnarray}
R_W =
\frac{1}{3}
[erf(A_+)+erf(A_-)]\nonumber \\
Q_W =
1-\frac{1}{3}erf(A_0)+\frac{1}{3}erf(A_-)-\frac{1}{3}erf(A_+)\nonumber \\
\label{mapD}
\end{eqnarray}  
were $A_{\pm}=\frac{{\rho_J}/{\sqrt{T}} \pm {\lambda_1}/{\sqrt{Q_J}}} 
{\sqrt{2 (1-\rho_J^2)}}$ and $A_{0}=\frac{\lambda_1} 
{\sqrt{2 Q_J (1-\rho_J^2)}}$. 

From Eq. \ref{mapD} one can verify that at the limit $\alpha \rightarrow 
\infty$ when the continuous order parameters achieve a perfect learning, 
$\rho_J \rightarrow 1$, the discrete order parameters  achieve perfect 
learning as well, $R_W \rightarrow 2/3$, $Q_W \rightarrow 2/3$ and  
$\rho_W \rightarrow 1$ given that the  positive quantity, 
$\lambda_1$, is smaller than $\lambda_1<\sqrt{Q_J/T}$.

In general, in  order that the digital student will gain perfect learning
it is necessary that the relation $\sqrt{Q_J/T} (\l-1)<\lambda_l<\sqrt{Q_J/T}
 \l$ holds for any  positive $\l$.
Note that the  interpretation of the above constraint is that in the vicinity 
of perfect learning the precursor might be focused around any set of discrete 
symmetric values, but not necessarily the ones that the clipped student has.

One of the conclusions concerning $\lambda_\l$ is that the law according 
which $\epsilon_g$ decays is independent of the exact value of the limit 
value, $\lambda_{\l}$.
It depends only on the ruler (binary/continuous), the specific strategy 
of learning (on-line/off-line) and the learning algorithm one uses.
In the following we analyze all these variations.

\section{Binary output}

\subsection{On-line Learning}

In an on-line learning scenario one can write equations of motion that 
determine the development of the order parameters as a function of $\alpha$.
The rate of convergence depends on the rule, F (Eq. \ref{rule}) and the 
learning algorithm that one uses f (Eq. \ref{update}). 
Fine tunes are made by choosing the  learning rate,
$\eta$.

We analyze learning procedure in the case of binary rule,
\begin{equation} 
S=sign(x),
\end{equation}
where $x$ is the local field and  
the generalization error as a function of $\rho$ is known to be 
\begin{equation}
\epsilon_g =\frac{1}{\pi}
\cos^{-1}(\rho)
\label{eg}
\end{equation}

Although it was shown that using the ``expected stability'' algorithm 
that maximizes the generalization gain per example leads to an upper 
bound for the generalization ability, \cite{Caticha}, 
we choose to concentrate on the so-called AdaTron or relaxation 
learning algorithm. This latter algorithm for zero stability, 
$\kappa=0$, performs comparably well and unlike the ``expected stability'' 
algorithm  does not require additional computations in the student network 
besides the updating of its weights, and the analysis is simpler as well
 \cite{PercBiehl}.

The convergence to perfect learning depends on the learning rate, 
if it is too large perfect generalization  becomes impossible. 
The transition from learnable situation to unlearnable occurs at  $\eta_c$.
In the following, in order to simplify the analysis; 
we choose a fixed learning rate, $\eta=1$, 
which is below  $\eta_c$  in all scenarios.

We update the artificial continuous weight vector, $\vec{J}$. 
The updating is made as in Eq. \ref{update} 
according to the following learning rule: 
\begin{equation}
J_i^{\mu+1} =J_i^{\mu}
 -\frac{\eta}{\sqrt{N}}
(\frac{\vec{J}^{\mu}\cdot\vec{\xi}^{\mu}}{\sqrt{N}}) 
\xi^{\mu}_i
\theta(-\frac{\vec{J}^{\mu}\cdot\vec{\xi}^{\mu}}{\sqrt{N}} S^\mu)
\end{equation}
The equations for the order parameters  with $\eta=1$ are,
\begin{eqnarray}
\frac{d\rho_J}{d\alpha} =
-\frac{\rho_J}{2 \pi}\cos^{-1}(\rho_J)
+\frac{1}{\pi}(1-\frac{\rho_J^2}{2})\sqrt{1-\rho_J^2}, \nonumber \\
\frac{dQ_J}{d\alpha} =\frac{Q_J}{\pi}
[\rho_J \sqrt{1-\rho_J^2}
-\cos^{-1}(\rho_J)]. \nonumber \\
\label{binary}
\end{eqnarray}
In the limit  $\alpha \rightarrow \infty$,
one obtains a power law that describes the convergence of $\rho_J$ and $Q_J$,
\begin{eqnarray}
\rho_J  \sim 1-2(\frac{3 \pi}{4})^2\frac{1}{\alpha^2}
 \nonumber \\
Q_J \sim Q_0 (1-\pi^2 (\frac{3 }{4})^3\frac{1}{\alpha^2})
\nonumber \\
\label{converge}
\end{eqnarray} 
Note: since we have a binary output unit, perfect learning is gained as soon 
as the angle between the vectors goes to zero independent of the 
student's norm.

The solution of Eq. \ref{binary} only describes the development of 
the continuous  perceptron's overlaps.
The next step is mapping the continuous precursor to the clipped one.
Since in the case of binary ruler the student's norm converges to some 
unknown value, one way of choosing  $\lambda_\l$ is simply 
 ``half the way'' between the constrained values, i.e.
 $\lambda_{-L}= \lambda_{L+1}=infty$ and otherwise
\begin{equation} 
\lambda_\l=\frac{1}{L}(\l-\frac{1}{2}) \sqrt{\frac{Q_J}{T}}.
\label{lim}
\end{equation}

The development of the order parameter $\rho_J$, 
is independent of the norm $Q_J$. Using a limit set that scales with 
$\sqrt{Q_J}$, Eq. \ref{lim}, ends up in $\rho_W$ which depends on $\rho_J$
but does not depend on  $Q_J$. 
Hence, plugging into it  Eq. \ref{converge}, 
one can find the the asymptotic behavior of the generalization error,   
Eq. \ref{eg} in the limit $\alpha \rightarrow \infty$  

\begin{equation}
\epsilon_g \propto 
\frac{\exp{(-K(\lambda)\alpha^2)}}{\alpha^{\frac{1}{2}}},
\label{limeg}
\end{equation}
where $K(\lambda)=\min|\lambda_\l-\frac{\l}{L} \sqrt{Q_J/T}|$.

We exemplify the aforementioned discussion in the diluted Ising perceptron.
We  use the  limits as in \ref{limits} 
and assume $\lambda_1=c \sqrt{Q_J/T}$.
In that case 
\begin{equation}
\rho_W =
\frac{erf(a_+)+erf(a_-)}
{ \sqrt{T} \sqrt{9-3erf(a_0)-3erf(a_+)+3erf(a_-)}},
\label{Maprho}
\end{equation}
where $a_\pm=\frac{\rho_J \pm c}{\sqrt{2 T(1-\rho_J^2)}}$ and 
 $a_0=\frac{c}{\sqrt{2 T(1-\rho_J^2)}}$.  
In the limit of large $\alpha$ one  finds
\begin{equation}
\epsilon_g \propto 
\frac{\exp{(-b_c \alpha^2)}}{\alpha^{\frac{1}{2}}},
\label{Dec}
\end{equation}
where for $c \geq 1/2$ $b_c=\frac{c^2}{3 \pi^2}$ and otherwise
$b_c=\frac{(1-c)^2}{3 \pi^2}$.
One can see that choosing $c=1/2$ results in a fastest decay of the 
generalization error.

\begin{figure}[t]
\begin{center}
\setlength{\unitlength}{3pt}
\begin{picture}(140,92)(0,0)
\put(0,0){\makebox(140,92)
          {\includegraphics{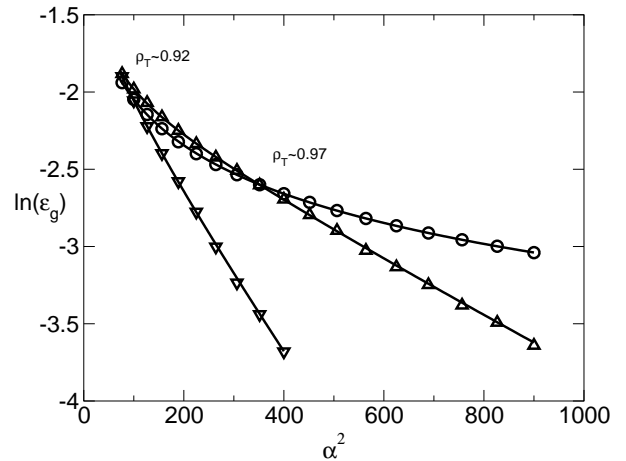}}}
\caption{ Simulation results of  $ln(\epsilon_g)$ of the continuous precursor 
($\circ$) and of the clipped vector  vs. $\alpha^2$. 
The clipping is made according to the mapping in  \ref{limits},
where  the results are for $\lambda_1=0.5\sqrt{{Q_J}/{T}}$ 
($\bigtriangledown$) and $\lambda_1=0.3\sqrt{{Q_J}/{T}}$ ($\bigtriangleup$).
error bars are smaller then symbols.
Solid lines are the numerical integrals (Eq. \ref{binary}).
 $\rho_T$ refers to the point at which a transition occures  between  
a superior performance by continuous/clipped perceptron, (see text).} 
\label{learn}
\end{picture}
\end{center}
\end{figure}
 \vspace{-1mm}

The analytical results are compared with simulations
on a teacher of the type of the  diluted Ising perceptron
with the following parameters;
$\lambda=0.5 \sqrt{Q_J/T}$ and $\lambda=0.3 \sqrt{Q_J/T}$, 
see Figure \ref{learn}.
The initial conditions for the continuous student weight vector are 
$Q_J(\alpha=0)=T=2/3$ and  $R_J(\alpha=0)=0$. 
The weight components were drawn out of a Gaussian distribution.
We used $\eta=1$, $N=3000$ and each point was averaged over $50$ samples.
One can see in Figure \ref{learn} that the analytical results give by Eq.
 \ref{Maprho} and Eq. \ref{Dec} are in agreement with simulations.

One can see that the super-exponentially decay is independent of the 
accurate value of $\lambda$. 
However, two important parameters do depend on the exact choice of $\lambda$.
One is the decay rate, the factor $K(\lambda)$ in the 
large $\alpha$ limit. One can see, for instance, that the optimal limit, 
$\lambda=0.5$ results in a faster decay than the limit $\lambda=0.3$.
The second is the exact $\alpha$ or the exact value of $\rho_J$ 
at which the clipped version gives a better result than the continuous one.
We named this value as  $\rho_T$. 
For $\rho_J<\rho_T$ the clipping lowers the overlap  $\rho_J$ since  the 
learning solution does not contain enough information about the real 
direction of the teacher, $\vec{W}^T$, so that clipping only leads the 
solution to forget a
little about the learned pattern without bringing it closer to the exact
solution. In the other region, when $\rho_J>\rho_T$, clipping becomes 
efficient because the learning solution is near the exact one.
The numerical results of  $\rho_T$ according
 to the mapping, (Eq. \ref{Maprho}), are 
 $\rho_T \sim 0.92$ for  
$\lambda_1=0.5\sqrt{{Q_J}/{T}}$ and $\rho_T \sim 0.97$ for
$\lambda_1=0.3\sqrt{{Q_J}/{T}}$, see Figure \ref{learn}.

\subsection{Clipped -Hebbian Learning}

Ising perceptron, diluted Ising perceptron and all the binary units
that are confined to discrete values exhibit a phase transition 
\cite{MeirFontanary,Derr,Kinzel}.
This known result was hard to achieved by a practical algorithm.
One way to gain a perfect learning is to include the information
of all the patterns simultaneously in the weights  by using
 the Hebb learning procedure, Eq. \ref{Hebb}.
Such a learning will end up in a discrete student only in the limit 
$\alpha \rightarrow \infty$. 
The decay of the generalization error in that case is known, 
since it is exactly the same as having a continuous teacher, 
$\epsilon_g \propto 1/\sqrt{\alpha}$ \cite{Vallet,ValletGailton}.
In such a way the knowledge of having digital values is not used, one has 
a continuous student that happens to realize, after learning, that the values
are constrained to a finite depth.

\begin{figure}[t]
\begin{center}
\setlength{\unitlength}{3pt}
\begin{picture}(140,92)(0,0)
\put(0,0){\makebox(140,86)
          {\includegraphics{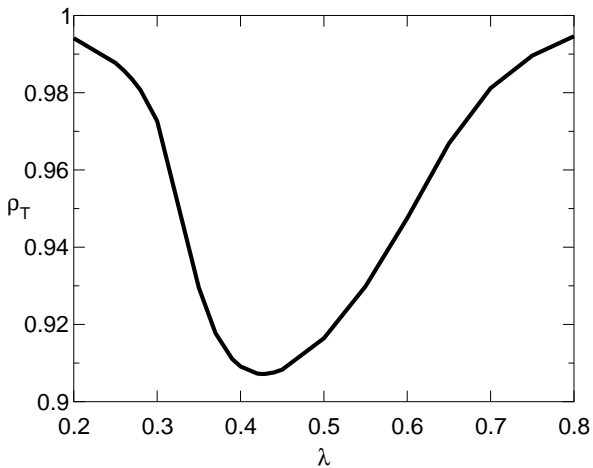}}}
\caption{ Analytical results of $\rho_T$,  as a function of the limit 
($\lambda_1$) in the  diluted  Ising case. 
$\rho_T$ stands for the continuous overlap value at which below/above  
it, a better generalization  is achieved by the continuous/clipped perceptron.} 
\label{rhoT}
\end{picture}
\end{center}
\end{figure}
 \vspace{-1mm}

The above mentioned procedure describes a way of using two vectors.
A continuous one, which is evaluated according to the Hebb rule and
a discrete student, obtained by clipping the continuous precursor  
according to Eq. \ref{map}. The latter mapping results in a better 
generalization error for large enough $\alpha$.

We take for example the diluted Ising case.
Given that the continuous student is normalized to be the same as the teacher 
one, $Q_J=T=2/3$, one can find the exact point,  $\rho_T$
at which the clipped method 
results in a better generalization. 
This value depends on the limit one chooses. One can see that the limit 
$\lambda_1$, that results in a  better generalization of the clipped version 
in smaller $\rho$, is $\lambda_1 \sim 0.43 \sqrt{Q_J/T}$, 
see  Figure \ref{rhoT}.
One might anticipate to get as a result $\lambda_1 \sim 0.5$, that  was found 
to optimize the decay as $\alpha \rightarrow \infty$.
However, the above value is determined by the distribution of the continuous 
weights in the beginning of the learning process, small $\alpha$. 
In this regime  the distribution of the weights  is close to  a Gaussian, 
and its tail influences the value of $\lambda_1$. 
The analysis above indicates that choosing  $\alpha $--dependent limits, 
$\lambda(\alpha)$ in this specific case might perform an even better 
generalization curve.

To conclude, the benefit from the clipping is evident only after the 
Hebb solution is near the exact one, after gaining  large $\rho$.
For optimizing the learning time, choosing the limits should be done 
cautiously. If the aim of the learning is to minimize the generalization 
error at the very end of the procedure, after a long learning process,
than the best choice for the limit will be the ``half the way'' method,
 Eq. \ref{lim}. However, to minimize the  generalization error  
for a  given finite $\alpha$, the best value  might be around 
$\lambda_1 \sim 0.425 \sqrt{Q_J/T}$.
These results suggest that it is possible to optimize the generalization 
error of the clipped perceptron by the choice of  a dynamical  
$\lambda_1=\lambda_1(\alpha)$.

\section{Large synaptic depth }

In this section we examine the crossover of the generalization error in the 
presence of  continuous weights as we increase the synaptic depth.
As long as the synaptic depth $L < O(\sqrt{N})$,  the generalization error
still vanishes  super-exponentially, Eq. \ref{limeg},  where the pre-factor 
decreases with $L$.  For $L \ge O(\sqrt{N})$ the learning is characterized 
by the features of spherical constrained learning.

A first step towards the continuous case limit is to find out
the change of the decay of the generalization error as a function of L.
We focus on the binary unit in the on-line scenario. The analytic 
tractability of this model enables a profound study of the influence of the 
synaptic depth over the learning features.

In the last model the generalization decays super-exponentionall,
$\epsilon_g \sim \exp(-K \alpha^2)$, (see Eq. \ref{limeg}).
The factor $K$ depends on the limits one chooses, $\lambda_l$. 
Hence, in order to keep on consistency, we use the abovementioned limits, 
(Eq. \ref{lim}), in the different depths cases.
We should emphasize at this stage that only one out of many 
super-exponentionall terms that arise from the asymptotic expansion of all the 
error functions (Eq. \ref{RQ}), was kept (Eq. \ref{limeg}). 
As soon as the deviations between different factors in the 
exponent are too small, one has to integrate all the terms together 
instead of neglecting all but one. Such a procedure results in a different 
type of decay, a power law instead of a  super-exponentionall decay.

\begin{figure}[t]
\begin{center}
\setlength{\unitlength}{3pt}
\begin{picture}(140,95)(0,0)
\put(0,0){\makebox(140,95)
          {\includegraphics{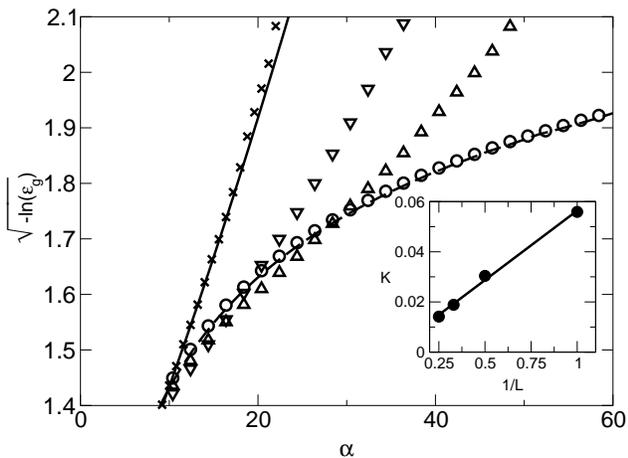}}}
\caption{ 
Simulation results of  $\sqrt{-ln(\epsilon_g)}$ in the case of 
$L=1$, (diluted Ising) ($\times$), 
$L=2$ ($\bigtriangledown$), 
$L=3$ ($\bigtriangleup$) and 
$L=157$ ($\bigcirc$) versus  $\alpha$.
The analytical results obtained by the numerical integration of Eq. 
\ref{binary} and Eq. \ref{Maprho} is presented for the ISing case (solid line).
The dashed line is the analytical curve for  $\sqrt{-ln(\epsilon_g^J)}$, 
were $\epsilon_g^J$ is the generalization error of the {\it continuous} 
student.
Inset: The dependece of the prefactor $K(L)$ on the depth $L$, in 
Eq. \ref{limeg}.
Solid line is the least squered fit, $K=0.06/L$.} 
\label{L}
\end{picture}
\end{center}
\end{figure}

Analytical and simulations results of the generalization error
in varieties of synaptic depths are presented in Figure \ref{L}.
Simulations were carried out with $N=630$ and each point is averaged 
over $100$ samples.
The insert shows the estimated slope $K$, (Eq. \ref{limeg}), as a function 
of the depth  $L$. One can see that $K$ decreases linearly with  $1/L$.
The deviation from the analytically predicted 
interplay for large $\alpha$, $K \propto 1/L$, 
is probably due to  finite $N$ effects.

 \vspace{-1mm}

In the following we present argument supporting the statement that the 
generalization performance of finite depth machines  coincide with the 
performance of continuous machines as soon as $L \sim \sqrt{N}$.
This scaling is found by taking into account that:
(a) The difference between two available values is of order of $1/L$.
(b) The distribution of the continuous student values around the teacher's 
one is a Gaussian with a variance of $\sqrt{1-\rho_J^2}=1/\epsilon_g^J$, 
where $\epsilon_g^J$ is the generalization error of the  continuous student.
Having a learning procedure (in the continuous space) in a finite dimension 
results in a generalization error, 
$\epsilon_g^J$, which is different then the analytical predictions.
The variance is of order of $ \sqrt{1/N}$ \cite{SollichBarber}.
Hence, an estimation to the order of the lower value that $\epsilon_g^J$  
gets in a specific  run will be $ \sqrt{1/N}$.
As a  consequence, having a discrete machine of depth $L$  when 
\begin{equation}
\frac{1}{L} <<  \sqrt{1-\rho_J^2} \sim \epsilon_g^J \sim  \sqrt{\frac{1}{N}}
\label{relat}
\end{equation}   
or $L>> \sqrt{N}$,
gives the same results as those of continuous learning.
Note that Eq. \ref{relat} is consistent with the mathematical constraint that
was pointed out in section III when we discussed the continuous limit.
The simulations shows indeed that in the case of $L=157 >> \sqrt{N}$,
were $N=630$ the discrete vector's performance coincide with the analytical
learning curve of the {\it continuous} student.

It is worth pointing out that a similar result was found when analyzing the 
possibility of learning from a discrete teacher by a discrete student using 
a general updating rule, \cite{KinzelUrbanczik}.
The last analysis uses totally different argument results in the conclusion 
that only when the teacher's depth is of order $\sqrt{N}$, it is possible to 
learn the rule using an updating rule that depends on the discrete
weights, i.e. only  then it behaves  as if we have a continuous machine.

\section{Finite Systems - Perfect Learning}

The theoretical results presented in the  previous chapters  exhibit
the typical behavior of the generalization error and the order parameters.
The main result is the fast decay of the generalization error
of the clipped perceptron to zero, Eq \ref{limeg}.
In the case of teacher and student with continuous weights and finite $N$, 
the generalization error is always finite distance from zero, 
even in the asymptotic stage of the learning process. 
In contrast to the continuous case,  the learning of a perceptron with 
discrete weights and finite $N$   is characterized by a transition to  
perfect learning, as was found for  the Ising perceptron, \cite{Mi}.
Performing simulations in  that case results in a perfect
learning in some stage, since in the clipping version the 
student knows exactly the teacher's optional values.
Hence, the overlap becomes exactly one, $\rho_W=1$, 
and the generalization error becomes exactly zero as well, $\epsilon_g=0$.

In order to give an estimation to the number of steps needed for  getting 
perfect learning, $\alpha_f$, we use the following approximation valid in 
the $\alpha \rightarrow \infty$  regime, where
we can give an analytical approximation to the  interdependence
of $\rho_W$ and $\alpha$.
In addition, the minimal step before perfect learning is well defined: 
$\rho_W=1-2/(LN)$  or $\epsilon_g \sim \sqrt{1/(LN)}$.
Hence, we can find the interplay between  $\alpha$ and N.

\begin{figure}[t]
\begin{center}
\setlength{\unitlength}{3pt}
\begin{picture}(140,95)(0,0)
\put(0,0){\makebox(140,95)
          {\includegraphics{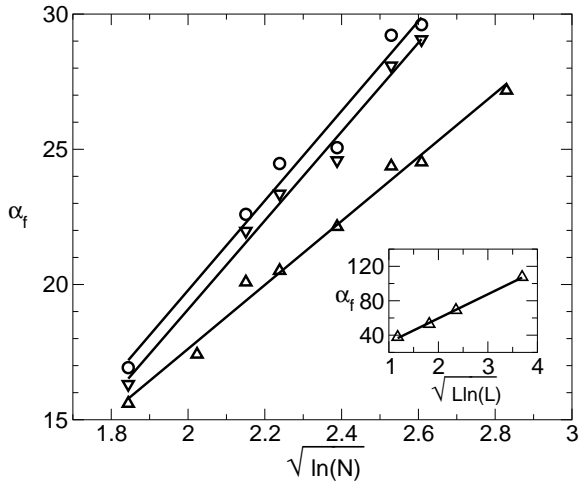}}}
\caption{ Simulation results of $\alpha_f$, the number of rescaled steps
necessary to achieve a  perfect learning vs. $\sqrt{\ln{N}}$.
 Simulations for diluted Ising perceptron, 
in the case of binary output unit, with 
$\lambda_1=0.4\sqrt{{Q_J}/{T}}$ ($\bigtriangledown$), 
$\lambda_1=0.5\sqrt{{Q_J}/{T}}$ ($\triangle$) and 
$\lambda_1=0.6\sqrt{{Q_J}/{T}}$ ($\circ$). 
 Solid lines correspond to  the linear fit of least
square error. 
Inset: Simulation results of $\alpha_f$ vs. $\sqrt{L \ln{L}}$
for $N=630$, $L=2,3,4,7$ and the limit values are chosen 
according to Eq. \ref{lim}.
Solid line is least squared fit.
}
\label{alphaf}
\end{picture}
\end{center}
\end{figure}
 \vspace{-1mm}

In the binary output perceptron  the generalization error 
falls down  super-exponentially, Eq. \ref{limeg}. 
Hence,  the  perfect learning is determined by 
\begin{equation}
\exp{(-K(\lambda,L)\alpha^2)}
\sim 
\sqrt{1/(LN)},
\label{neglect}
\end{equation} 
and since we found in the last chapter that $K$ decays linearly with $1/L$
we can derive $\alpha_f$ from the last equation, 
$\alpha_f \sim \sqrt{L \ln{ LN}}$.
This result indicates quantitatively that for any chosen limit, $\lambda_\l$, 
the number of  learning step necessary 
to achieve perfect learning is finite as long as $N$ and $L$ are finite.

Figure \ref{alphaf} presents results of  $\alpha_f$  obtained in simulations 
for the diluted Ising perceptron  with
$c=0.4$, $c=0.5$, and $c=0.6$, (Eq. \ref{Maprho}, \ref{Dec}).
Results  were averaged over  $M(N)$ training sets, were
values of $M(N)$  ranging from  $5000$ to $20$ in accordance to $N$ 
which is varied between $30$ and $9000$. 
To get results in lower dimension, $N$, we averaged 
over a larger number of simulations, $M$.

One can see from the obtained values of  $\alpha_f(N,c)$ 
in Figure  \ref{alphaf}, that the last quantity is indeed linear in
 $\sqrt{\ln N}$.
Note that the obtained slope in Figure \ref{alphaf} for $c=0.4$ and $c=0.6$ is 
the same as it is expected since $b_c$ is symmetric around  $c=1/2$. 
In the inset, one can see that  $\alpha_f(L)$ in the case of $N=630$, indeed
increases linearly with $\sqrt{L \ln{L}}$. As $L \rightarrow \infty$ 
an infinite
number of examples are needed for perfect learning, there is a crossover to 
the spherical case as was discussed in the previous chapter.

Small deviations from a straight line in Figure  \ref{alphaf} are 
expected to be a consequence of the following approximations:
(a) We took as an analytical curve (Eq. \ref{neglect})
only the asymptotic function which is an expansion valid in infinite $\alpha$.
(b) We neglected the polynomial corrections in Eq. \ref{neglect} such as 
$1/\sqrt{\alpha}$.
(c) We derived Eq. \ref{neglect} from the analytical calculation 
of $\rho_J(\alpha)$.  The latter quantity itself is influenced by finite size 
effects. 
Extensive numerical simulations show that the corrections are linear in 
$1/N$ \cite{DerrGriff,BuhotMorenoGordon,SollichBarber} 
and hence they are negligible after clipping and getting $\rho_W$ 
(As in Eq. \ref{Map}).

As was shown in previous chapters, $c=0.5$ gives the best performance in the 
asymptotic learning procedure,  lower  $\alpha_f$ for all N, 
and it is confirmed in our simulations, Figure \ref{alphaf} .
In the thermodynamic limit  $N \rightarrow \infty$,   
$\alpha_f \rightarrow \infty$ as expected.

\section{Continuous Unit}

We now study the case of continuous output perceptrons with finite depth.
As long as one uses a continuous activation function, 
the generalization error decreases exponentially, 
(see for instance  \cite{SollaSaad,RosenZviBiehlKanter,BiehlSchwarze}). 
In order to learn a rule which is defined by a finite depth vector,  
we used a spherical vector for the student weight vector, $\vec{J}$, 
and clipped it in order to have a digital student weight vector $\vec{W^S}$. 
The updating of the spherical student weight vector is done according to the 
gradient descent method as usual:
\begin{equation}
\vec{J}^{\mu+1}
=\vec{J}^{\mu}-\frac{\eta}{\sqrt{N}}
\nabla_{\vec{J}} \ \epsilon(\vec{J}^\mu,\vec{\xi}^\mu)
\end{equation} 
The error $\epsilon(\vec{J}^\mu,\vec{\xi}^\mu)$ measures the deviation 
of the student from the teacher's output for a particular input $\vec{\xi}$.
The generalization error of a student is defined as the averaged error
\begin{equation}
\epsilon_g=<\frac{1}{2}[S(\vec{J},\vec{\xi})-S(\vec{W}^T,
\vec{\xi})]^2>_{\vec{\xi}}.
\label{AvEg}
\end{equation}

Since the learning features  of all kinds of the continuous transfer functions 
are more or less the same, we chose to concentrate in 
the ``sin'' activation function 
\begin{equation}
S=\sin(kx), 
\end{equation}
The periodic activation function, $\sin$, was found to be learnable given 
that the period k is small enough \cite{RosenZviBiehlKanter}.
In the following we will simplify our analysis by taking k=1
and the  learning rate $\eta=1$.
Since the learning curves of the continuous version are the same as if there
 was a rule defined by a continuous teacher, (having the finite depth 
limitation is merely a special case of the spherical constraint), 
and the learning rate we chose is small enough we  find
that perfect learning is an attractive fixed point in both scenarios.

Linearizing the equations of motion around those fixed points 
results in the following form (which holds for all the continuous transfer
functions):
\begin{eqnarray}
R_J=1-\frac{c_1}{\det{\matrix{V}}} V_{22} \exp{(\gamma_1 \alpha)}
+\frac{c_2}{\det{\matrix{V}}} V_{12} \exp{(\gamma_2 \alpha)} \nonumber \\
Q_J=1+\frac{c_1}{\det{\matrix{V}}} V_{21} \exp{(\gamma_1 \alpha)}
-\frac{c_2}{\det{\matrix{V}}} V_{11} \exp{(\gamma_2 \alpha)} \nonumber \\
\end{eqnarray}
The two eigenvalues of $\matrix{V}$, $\gamma_1$, $\gamma_2$, 
are both negative. The constants $c_1$, $c_2$  are 
determined from the numerical solution of the equations of motion. 

In order to get a description of the discrete learning one has to use
the mapping relations as in Eq. \ref{map}.
The generalization error of the finite depth student 
directly depends on the order parameters, as can be found by taking the 
average over the local fields distribution, Eq. \ref{AvEg}.
The general result of this calculation at the $\alpha \rightarrow \infty$
regime is  
\begin{equation}
\epsilon_g \sim
\exp{(-C_0e^{|K| \alpha})},
\end{equation} 
were $K$ and $C_0$ depend only on the learning rate ,$\eta$, the limits
one chose, $\lambda_\l$ and the specific activation function.
In the following we examine this result in the diluted Ising case.

\begin{figure}[t]
\begin{center}
\setlength{\unitlength}{3pt}
\begin{picture}(140,92)(0,0)
\put(0,0){\makebox(140,92)
          {\includegraphics{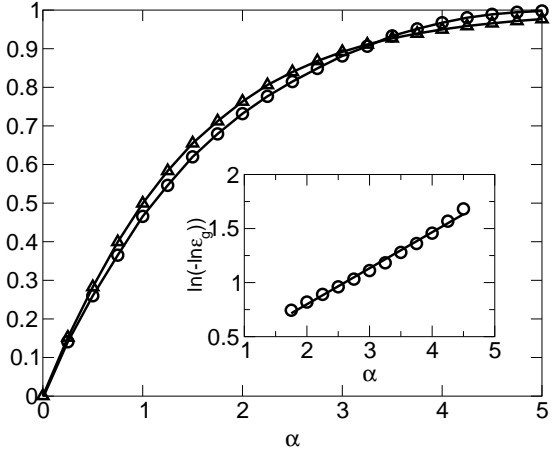}}}
\caption{ Simulation results of  $\rho_J$ ($\bigtriangleup$) and $\rho_W$
($\circ$) vs. $\alpha$ in the diluted Ising case. 
Solid lines are the numerical integrals (Eq. \ref{mapD}, \ref{singlediff}).
Inset: $\ln(-\ln(\epsilon_g))$ vs. $\alpha$ obtained in simulations ($\circ$)
with $N=3000$. Solid line is least squared linear fit, the slope was found to 
be 0.33.} 
\label{learnC}
\end{picture}
\end{center}
\end{figure}
 \vspace{-1mm}

We performed simulations in the diluted Ising case, when the transfer 
function is $\sin$. The development of the continuous order parameters
in that case is described by the following equations of motion,
\begin{eqnarray} 
   \frac{dR_J}{d\alpha} & = & \frac{1}{2} \left[\!(R_J\!+\!1)
          D_{+}\!-\!2R_Je^{-2Q_J}\!-\!(R_J\!-\!1) D_{-}\!\right]  \nonumber \\
   \frac{dQ_J}{d\alpha} & = &  \left[(R_J\!+\!Q_J)D_{+}\!-\!2Q_Je^{-2Q_J}
  \!-\!(Q_J\!-\!R_J)D_{-}\right]   \nonumber  \\
  &+& \!\! \frac{1}{8}  
  \left[
   2 \left( e^{-2 Q_J} - e^{-2} 
   -  E_{-} + D_{+} \right)  
   \right. 
 \nonumber \\
   &+ &\!\! 3\!-\!D_{-}^4\!-\! 2 D_{-}\! 
   \left. +\!\left( 2 E_{+} - e^{-8 Q_J} - D_{+}^4 \right) 
    \right] \! \label{singlediff} 
  \end{eqnarray}
with $D_{\pm} = e^{-(1+Q_J \pm 2R_J)/2}$ and $E_{\pm} = e^{-(1+9Q_J \pm 6R_J)
 /2} $.  
As $\alpha \rightarrow \infty$, one gets two eigenvalues,
$\gamma_1 \sim -0.30$, $\gamma_2 \sim -0.69$.
Using Eq. \ref{mapD}, rescaling $R_W$ and $Q_W$ by the teacher's 
norm, $2/3$, and taking the limit value, $\lambda$, to be the one that yields 
the faster decay at the large $\alpha$ regime, 
$\lambda=0.5 \sqrt{Q_J/T}$.  Collecting everything we have
\begin{eqnarray}
R_W \sim 1-\frac{\exp{(-0.15 \alpha)}}{2 \sqrt{\pi} K_1}
\exp{(-K_1^2e^{0.30\alpha})}
\nonumber \\
Q_W \sim 1+\frac{\exp{(-0.15 \alpha)}}{ \sqrt{\pi} K_1}
\exp{(-K_1^2e^{0.30\alpha})},
\nonumber \\
\label{AsymD}
\end{eqnarray}
where K is determined by the initial conditions.
The generalization error as a function of the discrete parameters is 
\begin{equation}
\epsilon_g=\!\frac{1}{2} \left[1\!-\!d_{-}\!+\!  
   d_{+}\!-\!\frac{1}{2}\left(e^{-2 Q_W}\!+\!e^{-2 }\!\right)\right]
\label{singlegenerror}
\end{equation}
with $d_{\pm} = e^{-(1+Q_W \pm 2R_W)/2}$.
Expanding the last equation around $R_W \rightarrow 1$ and $Q_W \rightarrow 1$,
 we obtain that the generalization error decays  very fast, 
$\epsilon_g \sim \exp{(-K_1^2e^{0.30\alpha})} $. 
 
We ran simulations with $N=3000$ and averaged over $10$ samples.
In Figure \ref{learnC} the development of the discrete as
well as the continuous order parameters as a 
function of  $\alpha$ are presented.
The solid lines are the analytical numerical integrals of Eq. \ref{singlediff}.
 Note, the transition in this scenario from a poor 
generalization of the clipped version comparatively to that of the continuous
one,  to a situation in which the clipped version has a better performance, 
occurs in the same $\rho_T \sim 0.92$ as in the binary unit.
This quantity is related to the clipping rule and it is independent of the
 specific transfer function one tries to learn.
 
The inset of Figure \ref{learnC} shows the unique decay of the 
generalization error, in order to get linear line we plotted 
$\ln{(-\ln{\epsilon_g})}$ as a function of $\alpha$. 
According to the above analysis the slope should be 0.30 and we obtained
in simulations $0.33 \pm 0.01$. 
It is  in  good agreement, considering the fact 
that we are dealing with an approximation which is valid only in the
$\alpha \rightarrow \infty$ and simulations results are in finite $\alpha$.
The generalization error of the clipped version for
  larger $\alpha$ ($\alpha>7$ in our case) 
gives better results than those predicted by the analysis, 
its values are exactly zero due to finite size effects discussed 
in chapter V.
 
Following the same arguments used in order to find an estimation
to the number of examples needed for gaining perfect learning, one finds that 
in the case of continuous output $\alpha_f \sim \ln{(\ln{N})}$.
It is obvious from the analytical calculations and the simulations
above that clipping a continuous vector in order to learn a finite depth 
teacher results in an extremely fast learning.
The learning in finite dimension is characterized by  $\alpha_f$, above which 
one gets perfect learning of the discrete vector.
All those unique characteristics of the discrete learning disappears 
as soon as the weight depth is of order of $\sqrt{N}$ as was found in chapter 
VI.

\section{Conclusions}

In this paper, we presented an analysis of the simplest neural network,
the perceptron, that learns from examples given by another perceptron, 
the teacher, which is confined to a discrete space. 
In fact, we used two students, a continuous precursor and its
clipped version.

We analyzed the new set of order parameters arising from the clipping
 method. We discussed the issue of how to clip and what set of limits, 
$\lambda_\l$, is the best choice.
We found that it depends specifically on the kind of optimization one imposes.
We showed that after reaching some overlap, $\rho_T$, a transition occurs
 and the clipped version results in a better performance then the non-clipped 
one. If one is interested in optimizing the learning in the sense of getting 
a better performance as soon as possible, then the  minimizing  $\rho_T$
limits are the ones needed.
However, if by optimizing one tries to get the fastest decrease possible in
the $\alpha \rightarrow \infty$ regime then the best choice is 'half the way',
in-between the values. As we mentioned before, it is possible to have a 
dynamic set of values that interpolates during the learning process between
both values. We left this issue out of the scope of this paper.

As one can see from the definitions in Eq. \ref{zero}, it is only natural to 
choose the continuous weight vector not to  be the one which is constrained 
to a hypersphere but a vector which is constrained to  
a hypercube space. It was shown that in the case of storing random patterns
pre-training a continuous student whose weight vectors constrained to the
 volume of a hypercube results in a better performance \cite{BoutenReimers}. 
It remains as an open question what is the quantitative benefit that one 
can gain in a learning procedure by using the cubical constrained and 
 if a learning strategy could be designed which fulfills this constraint.

We studied the case of a very large $L$ and show a scaling relation between 
$L$ and $N$ arises from the analysis.
For $L \sim  O(\sqrt{N})$ the learning curve is the one that is 
typical  to the continuous case.
However, it should remain clear that  learning is the same as 
having a continuous student unless
$\alpha \rightarrow \infty$, $\rho_J \rightarrow 1$.
In that regime the fast decay that 
characterizes the clipped learning appears.
All digital computers actually correspond to such a
situation, where all available properties have a finite representation. 
The machine is using some kind of clipping by rounding  the numbers.
The differences, as predicted here, can be significant only in the 
 $\alpha \rightarrow \infty$ regime or small depth. 
Visualizing them is usually impossible since they are smaller than 
the measurements scale.

\acknowledgments{}
Discussions with W. Kinzel and M. Biehl are acknowledged.
We would like to thank M. Biehl for carefully reading of the manuscript.
The research is supported by the German Israel foundation and the Israel 
Academy of Science.

\clearpage

\end{document}